\documentclass{ws-rv9x6}  

\begin{document}

\setcounter{chapter}{0}

\chapter{THE PARADIGM OF INFLATION}

\markboth{J. Garc\'{\i}a-Bellido}{The Paradigm of Inflation}

\author{J. Garc\'{\i}a-Bellido}

\address{Departmento de F\'{\i}sica Te\'orica,
Universidad Aut\'onoma de Madrid\\
Cantoblanco, 28049 Madrid, Spain.
E-mail: juan.garciabellido@uam.es}

\begin{abstract}
The standard model of cosmology is based on the hot Big Bang theory
and the inflationary paradigm. Recent precise observations of the
temperature and polarization anisotropies in the cosmic microwave
background and the matter distribution in large scale structures like
galaxies and clusters confirm the general paradigm and put severe
constrains on variations of this simple idea. In this essay I will
discuss the epistemological foundations of such a paradigm and
speculate on its possible realisation within a more fundamental
theory.
\end{abstract}

\section{Introduction}   

Modern cosmology stands today on a firm theoretical framework, based
on general relativity and the hot Big Bang theory, which describes
with great precision the evolution of the universe from the first
fraction of a second to our present epoch. However, this impressive
framework was unable to explain the flatness and homogeneity of space,
nor the origin of the matter and structures seen in the universe
today. The advent of the inflationary paradigm\cite{1,3,5} in the
1980's provided a dynamical mechanism for the generation of both
matter and structure in an otherwise flat and homogeneous
universe.\cite{4} Its predictions are confirmed today with great
accuracy, thanks to a true revolution in cosmological observations,
coming mainly from measurements of the anisotropies\cite{6,7} in the
cosmic microwave background (CMB) and the distribution of matter in
large scale structures (LSS) like galaxies and clusters of
galaxies.\cite{8,9}

Nowadays, the hot Big Bang theory and the inflationary paradigm
constitute the basis of our standard model of cosmology. However,
while the theory of the Big Bang is well established, there is no
theory of inflation yet. At the moment, most people consider inflation
as a successful idea, a paradigm, realised in specific models with
concrete predictions. I will give here an overview of the present
theoretical and observational status of the idea of inflation, within
the most general framework, trying to be as much model-independent as
possible.

I will first present an epistemological analogy that will help put
into context the successes of inflation and the search for a
fundamental theory of inflation. I will discuss the assumptions that
are used in the construction of the most general models of inflation,
with their main caveats. I will then explore its main consequences and
predictions, both theoretical and phenomenological.  One may think
that inflation is such a basic idea that one cannot rule it
out. However, I will describe ways in which one could discard not only
specific models of inflation but the whole paradigm. To end this broad
view of inflation, it might be useful also to state clearly some minor
criticisms to the idea of inflation, as well as dispell some false
claims that appear in the literature. Finally, although inflation
defines a general framework in which most cosmological questions can
be addressed, it is certainly not the panacea of all problems in
cosmology, so it is worth mentioning those questions inflation cannot
answer, at least in its present realisation. At the end, I will give a
personal perspective on what I think should be the basic ingredients
of a complete theory of inflation.

\section{An epistemological analogy}   

The importance and outreach of a paradigm or a theory is best put into
perspective by comparing it with other previous ideas whose
epistemology is well known to most researchers in the field. The idea
of inflation has sometimes been compared with the gauge principle of
the standard model of particle physics. Like the inflationary
paradigm, the gauge principle is realised in a specific model (based
on a concrete choice of gauge group) with specific predictions that
depend on a large number of fine-tuned parameters in order to agree
with observations. Fortunately, there is a significant number of
independent experiments that have fixed most of these parameters to
several decimal places in a consistent framework, at least up to the
energies reached with present accelerators. However, if we want to
extend the standard model to higher energies, we run into problems of
consistency. In order to solve those problems we need to introduce a
new theory with many new ingredients. This theory is still uncertain,
and will probably require new ideas based on experimental data beyond
the present energy scale. Perhaps with the future particle physics
colliders, at higher energies, we may find completely new and
unexpected phenomenology.  All of this can also be said of the
inflationary paradigm and its present realisation in terms of specific
models.

However, I feel this is not the best analogy for the inflationary
paradigm. What I find more appropriate as an epistemological analogue
is Newtonian mechanics. From our modern perspective, we can say that
Newton was lucky, he proposed a force law of gravitation without a
consistent theoretical framework of space and time, much to the
despair of his contemporaries, Leibniz and Berkeley.  However, he was
smart enough to state no mechanism for this force and propose his laws
of celestial mechanics with the statement: ``everything works as
if...''. What silenced the criticisms made by both physicists and
philosophers, like Mach and others, to the idea of absolute space and
time was the extraordinary power of Newtonian mechanics for predicting
planetary motion.

It was not until the advent of Einstein's theory of general relativity
that this extraordinary coincidence was dispelled, and put Newtonian
mechanics in the appropriate perspective as a concrete limit of a more
fundamental theory of space and time. I believe the inflationary
paradigm is nowadays in a similar situation with respect to a more
fundamental theory.  Present observations seem to be in complete
agreement with the main predictions of inflation, but we still do not
understand why it works so well. At a fundamental level, the idea of
inflation is outrageously simple, and yet contains all the ingredients
necessary to explain a large array of observations, many of which are
independent, as well as predicting a surprisingly rich phenomenology.
Moreover, most of its predictions were made much before we had any
means of making contact with observations.  Today, these observations
are confirming the general paradigm but still have not singled out any
particular model.

\section{Basic assumptions}

Before describing the successes of the idea of inflation let me begin
by stating clearly the basic assumptions that are used in the
construction of a generic inflationary model. Variations of this
generic model will be discussed below.

The inflationary paradigm assumes that gravity is described by a
classical field theory, usually taken to be general relativity in
4-dimensional space-time. This assumption is not essential, it can be
relaxed by including also a dilaton field, a scalar partner of the
graviton field, coupled to matter like in scalar-tensor theories, and
also by extending it to possibly extra compactified dimensions.

In order to explain the present homogeneity and flatness on large
scales, the only requirement is an early period of accelerated
expansion of the universe. In the context of general relativity, this
means a negative pressure as the source of gravity. The simplest
realisation is an approximately constant energy density, leading to
quasi-de Sitter (exponential) expansion of the early universe. It can
be parametrised in terms of an effective or fundamental scalar field,
called the inflaton field, whose nature is yet unknown. Other species
(matter and/or radiation) may be present, but will be diluted away
very quickly by the expansion. This inflaton field may evolve slowly
down its effective potential, or not. While an approximately constant
energy density seems to be required, a slow-roll field is only a
simplifying assumption.  Non-slow-roll models of inflation exist and
for the moment make predictions that are compatible with observations.

All matter fields, as well as fluctuations of the inflaton and the
gravitational field, are supposed to be quantum fields evolving in a
curved background. Quantum fluctuations of these fields start in the
vacuum at small scales. The expansion of the universe takes these
inflaton and metric fluctuations out of the de Sitter causal horizon and
produces classical metric perturbations at superhorizon scales. Later
on, during the radiation and matter eras, the causal horizon grows more
quickly than the size of the universe and fluctuations reenter again,
giving rise to radiation and matter perturbations as they fall into 
the potential wells of the metric.

\section{General consequences}

The main consequences of inflation can be classified in two groups:
those that affect the space-time background, and perturbations on this
background. Due to the tremendous expansion, inflation predicts that
spatial sections must be flat, that is, Euclidean, at least on our
local patch. It also predicts that these spatial sections must be
homogeneous and isotropic to a high degree. Both homogeneity and
flatness were unstable properties under the evolution equations of the
Big Bang theory and thus required highly fine-tuned initial
conditions.  Inflation eliminated the fine-tuning by postulating an
early period of acceleration of the universe. The rest is a
consequence of this simple hypothesis. For instance, if the universe
had a non-trivial topology before inflation, it was redshifted away by
the expansion, so that the topological cell today $-$ the geometrical
cell whose sides are nontrivially identified $-$ should be much larger
than the present Hubble volume. At the moment, this is in perfect 
agreement with observations of the CMB anisotropies.

Inflation not only explains in a dynamical way the observed
homogeneity and flatness, but also predicts the existence of
inhomogeneities and anisotropies arising from quantum fluctuations of
the inflaton and the metric, and possibly other scalar fields evolving
during inflation.  General relativity predicts that there must be six
physical degrees of freedom associated with a generic metric
perturbation: two scalar, two vector (the vorticity and shear fields)
and two tensor (the two polarizations of a gravitational wave)
components. The perturbed Einstein equations, together with the
inflaton field as matter source, imply that the two scalar metric
components are linearly dependent, while the vorticity field decays
very quickly during inflation. Therefore, we are left with only one
scalar (giving rise to curvature perturbations) and two tensor
components.

Due to quantum mechanics, superhorizon quantum fluctuations become
semi-classical metric perturbations. These perturbations have an
amplitude proportional to the approximately constant energy density
driving inflation and, therefore, give rise to almost scale-invariant
spectra of density perturbations and gravitational waves, that is,
fluctuations with approximately the same amplitude on all scales. The
density perturbations will later seed the structure we see in our
universe, as matter falls into them and evolves through gravitational
collapse. Deviations from exact scale invariance $-$ the tilt of the
spectrum $-$ characterise a particular model of inflation and thus allow
cosmologists to distinguish between them. In some cases, the tilt may
also depend on scale, which puts further constrains on models of
inflation. On the other hand, the tensor perturbations give rise to a
spectrum of gravitational waves that may eventually be detected by laser
interferometer gravitational wave observatories. Moreover, since quantum
fluctuations arise from essentially free fields during inflation, both
primordial spectra are expected to have a statistical distribution 
that is approximately gaussian.

If there were more than one scalar field evolving during inflation,
their quantum fluctuations may have given rise to both energy density
and entropy perturbations. The former arise from scalar curvature
perturbations, while the latter is responsible for isocurvature
perturbations. In some cases, this also implies a small
non-gaussianity in the spectrum, and more than one spectral tilt.
Unless required by observations, we will use Occam's razor and assume
that inflation is driven by a single scalar field.

\section{Observable predictions}

The main prediction of inflation is that spatial sections are
Euclidean, {\em i.e.} with negligible spatial curvature, at least
within our Hubble volume. This has been confirmed at the 2 percent
level by recent measurements of CMB anisotropies, as described
below. Inflation also predicts a homogeneous space-time background,
with inhomogeneities or anisotropies imprinted by small amplitude
quantum fluctuations giving rise to classical curvature
perturbations. The amplitude is not fixed by the general paradigm, but
can be computed precisely for a given model. In single-field
inflation, the amplitude of the curvature perturbation remains
constant outside the horizon and, therefore, its amplitude at reentry
is precisely given by that at horizon exit during inflation. Thus, a
generic prediction of inflation is a primordial spectrum of
scale-invariant curvature perturbations on superhorizon scales, that
will seed structure once they reenter the causal horizon and matter
falls into them, undergoing gravitational collapse. Such a primordial
spectrum was postulated in 1970, much before inflation, by Harrison
and Zeldovich in order to explain the present distribution of matter,
and I think it is extraordinary that such a simple idea as inflation
could also explain the origin of this peculiar spectrum in terms of
quantum fields in a curved background that, due to the expansion,
become classical metric perturbations.

In fact, curvature perturbations are also responsible for temperature
and polarization anisotropies in the CMB. As baryons fall into the
potential wells of these perturbations, they induce acoustic
oscillations in the plasma, with radiation pressure opposing
gravitational collapse. A perturbation of a given scale can only grow
after it enters the causal Hubble radius. Since all perturbations of
the same wavelength entered simultaneously, they must be in the same
phase of the acoustic oscillation at the time of decoupling. This
gives rise to spatially coherent oscillations that stand out as peaks
in the two-point angular correlation function. This pattern of
acoustic oscillations in the power spectrum of temperature
anisotropies is a characteristic signature of inflation, and has been
clearly seen in the CMB anisotropies by Boomerang and WMAP. Any
alternative theory of structure formation, like {\em e.g.} topological
defects, based on an active causal origin of matter perturbations,
could not explain the observed pattern of acoustic oscillations and
are therefore ruled out. It is extraordinary that the surprisingly
simple paradigm of inflation predicted the coherent oscillations much
before they could be measured. We are now using the detailed
properties of the observed spectrum to determine many cosmological
parameters.  For instance, the first acoustic peak in the power
spectrum appears at a scale that precisely corresponds to the
projected size of the acoustic horizon at decoupling, approximately
one degree in the sky today, and indicates that photons have travelled
since then in straight lines towards us.  This implies that the
universe is essentially flat, with deviations less than a few percent,
again as predicted by inflation.

Furthermore, according to inflation there will also be large scale
perturbations that are still outside the horizon at the time of
decoupling and therefore are not enhanced by the acoustic oscillations
of the plasma. These superhorizon perturbations are a specific
prediction of inflation, and are responsible for the so-called
Sachs-Wolfe plateau, arising from photons that are gravitationally
redshifted or blueshifted as they escape from or fall into these
primordial perturbations. The height of the plateau has direct
information about the amplitude of the primordial perturbations and the
flatness of the plateau about the tilt of the spectrum. Without an
acausal mechanism like inflation that can stretch fluctuations beyond
the horizon and imprint a primordial spectrum with superhorizon scales
at decoupling, we could not explain the Sachs-Wolfe plateau, observed by
the COsmic Background Explorer (COBE) satellite for the first time in
1992. To this plateau contributes not only the (scalar) density
perturbations, but also the (tensor) gravitational wave spectrum.

However, the fact that from our particular position in the sky we
cannot see more than a few realizations of the gaussian distribution
of large scale fluctuations $-$ an example of sample variance that here
goes by the name of cosmic variance $-$ means that the small tensor
contribution to the Sachs-Wolfe plateau may be hidden inside the
cosmic variance of the scalar component, and since gravitational waves
are quickly redshifted by the universe expansion after they enter the
Hubble radius, their imprint in the temperature power spectrum could
be unmeasurable after all. The primordial tensor spectrum is important
because it contains complementary information about the inflationary
dynamics, and, in particular, its amplitude depends only on the total
energy density driving inflation, which is still a great unknown.

Fortunately, the microwave background contains information not only
about the energy distribution of photons, a precise blackbody spectrum,
but also about their linear polarization. The polarization of the
microwave background is a vector field with a gradient and a curl
components (called E and B components in analogy with electromagnetism).
The amplitude of the polarization spectrum is directly predicted by the
size of the temperature fluctuations, since the linear polarization of
CMB photons can only arise via Thomson-scattering, from quadrupolar
anisotropies in orthogonal directions. This polarization is induced by
photons scattering off electrons in the last scattering surface, when
the universe was essentially neutral. As a consequence, the polarization
spectrum is an order of magnitude smaller than the temperature spectrum,
making it even more difficult to observe than the latter.

Only recently, DASI measured for the first time the polarization field
at the last scattering surface. However, it was WMAP, with much bigger
sky coverage and sensitivity, that measured the precise angular
dependence of the polarization field, and confirmed that the
temperature and the gradient (E) component of the polarization
spectrum were correlated. WMAP gave the multipole expansion of the TE
cross-correlation spectrum, with a pattern of peaks matching those
predicted long before by the simplest models of single-field
inflation. This is without any doubt one of the most important
arguments in favour of inflation. While the existence of the CMB
polarization spectrum is a generic prediction of Big Bang cosmology at
photon decoupling, it is extremely difficult to construct {\em ad hoc}
models of structure formation and yet be in agreement with this
pattern of acoustic oscillations in both the temperature and the
polarization power spectra.

Moreover, we still have the curl (B) component of the polarization
spectrum. This component can only arise from the (tensor)
gravitational wave primordial spectrum of perturbations coming from
inflation, and has not been detected yet. It may happen that the scale
of inflation is high enough for a significant amplitude of tensor
perturbations to generate a B polarization field at the level needed
for detection in the next generation of CMB satellites like Planck or
CMBpol. At this moment, there is a large experimental and theoretical
effort searching for ways to extract the rare primordial B-component
of the CMB, since it contains crucial information about the scale of
inflation that we may not be able to obtain otherwise.

Another interesting feature about CMB polarization is that is provides
a null test of inflation. We have seen that in the absence of
anisotropic stresses acting as sources during inflation, the vector
component of metric perturbations is negligible. While the scalar and
the tensor components are both symmetric under parity, the E- and the
B-components of polarization have opposite parity. Therefore, a robust
prediction of inflation is that there should be no correlation between
the scalar temperature (T) and the B-component of polarization, nor
between the E- and the B-components of polarization. This can be
summarised as $\langle{\rm BT}\rangle = \langle{\rm BE}\rangle = 0$,
and constitute an important test of inflation.  The other four
combinations, the three power spectra $\langle{\rm TT}\rangle,\
\langle{\rm EE}\rangle$ and $\langle{\rm BB}\rangle$, and the
$\langle{\rm TE}\rangle$ cross-correlation are predicted by inflation
to be non-zero and to possess a precise pattern of acoustic
oscillations.

Furthermore, CMB anisotropies are not the only probes of inflationary
predictions. The same primordial spectrum of curvature perturbations
responsible for temperature anisotropies gives rise, through
gravitational collapse of the primordial baryon and dark matter
distributions, to the large scale structures like galaxies, clusters and
superclusters. These perturbations are probed by the so-called matter
power spectrum, that is, the Fourier transform of the two-point
correlation function of luminous galaxies in clusters and superclusters,
all the way to the horizon in our Hubble volume. If light traces matter
then this power spectrum may have a precise relation to the primordial
spectrum of curvature perturbations, and thus to inflation and the CMB
anisotropies. The primordial adiabatic, gaussian and scale-invariant
spectrum of metric perturbations seeds, through gravitational collapse,
the measured matter power spectrum.  Galaxy surveys both in real and
redshift space have increased their covering volume at increasing rate
since the first CfA catalogs in the 1970s to the present 2dF Galaxy
Redshift Survey and the Sloan Digital Sky Survey, which cover a large
fraction of the universe up to redshifts or order $z=1$ for galaxies,
and order $z=3$ for quasars. These surveys have allowed cosmologists to
bring into sharp focus an image of the nearby universe to unprecedented
accuracy, confirming basic properties of the power spectrum, like
gaussianity and scale invariance, at least on large scales, where the
power spectrum has not been strongly distorted by non-linear
gravitational collapse.

In the near future, with the advent of the next generation of
gravitational wave observatories, it will be possible in principle to
detect the approximately scale-invariant spectrum of gravitational
waves generated during inflation, seen as a stochastic background in
time-correlated detectors across the Earth. However, this background
could remain undetectable, even by sensitive gravitational wave
observatories, if the scale of inflation is too low. The detection of
this new cosmic background, with specific signatures like a small
negative tilt, {\em i.e.} larger amplitudes at longer wavelengths, may
open the possibility of testing the idea of inflation. It happens that
in all single field models of inflation, there is a unique relation
between the two spectra (scalar and tensor) because the source of
these metric perturbations is a single function, the effective
inflaton potential. In particular, in models where the inflaton field
evolves slowly towards the end of inflation and reheating, the ratio
between the tensor and scalar amplitudes is uniquely related with the
tilt of the gravitational wave spectrum.  This is a prediction that no
other theory of structure formation could have imagined a priori, and
thus constitutes a test of the whole paradigm of inflation. If
confirmed by observations it would have tremendous impact on
inflation as a paradigm.

\section{The reheating of the universe}

The reheating epoch, in which the energy density of the inflaton decayed
into radiation, is also a universal feature of inflation. We know that
at some point inflation must end, at least in our local patch, since
today the universe contains the remnants of the radiation and matter
eras. This energy conversion epoch signals the beginning of the hot Big
Bang as we know it. However, the details of how it proceeds from the
quasi-stationary inflationary regime to the radiation dominated era is
rather model-dependent and requires some knowledge of the high energy
particle physics model in which inflation is embedded. In some models it
occurs in a perturbative way, as the quanta of the inflaton field decay
into other field quanta to which the inflaton couples. In other models
it occurs in an explosive non-perturbative way, producing a variety of
interesting phenomenology like non-thermal phase transitions, with
possibly the generation of topological defects, gravitational waves
and/or primordial magnetic fields. These features may leave their
imprint in the CMB anisotropies or in the stochastic gravitational wave
background or even in the primordial intergalactic magnetic fields.
Moreover, the far from equilibrium conversion of energy may help
generate the matter/antimatter asymmetry at reheating, in a way that is
now being explored both analytically and numerically.

Although none of these features are robust predictions of inflation,
once observed they could give crucial information about the theory of
high energy particle physics in which inflation is embedded, that is,
the couplings and masses of the other fields to which the inflaton
couples. For instance, the production of topological defects during
reheating after inflation is predicted in specific models of
inflation. Absence of any signatures from these cosmic defects can be
used to impose stringent constraints on those models, but do not rule
out other realisations of reheating.

What I think is outstanding of inflation is its ability to create such
a novel and rich phenomenology from such a simple hypothesis.
For inflation to work, very little has to be put in, and a lot seems
to comes out, and some of its predictions have far reaching consequences.

\section{Eternal inflation}

Although inflation was invented in order to alleviate the problem of
initial conditions of the Big Bang, it is not in itself a theory of
initial conditions. In fact, in order for inflation to proceed in the
first place it is assumed that it started in a single domain that was
sufficiently homogeneous across a causal horizon for the energy
density to be treated as approximately constant. Once inflation
started, this domain would occupy a large space-time volume due to the
rapid expansion and later give rise to the whole large scale structure
of space-time.\cite{2}

Until now, I have described the observable consequences of a generic
model of inflation, {\em i.e.} what is amenable to direct
experimentation in our Hubble volume, but there is also the
possibility of studying phenomena beyond our observable patch. The
inflationary paradigm has a mechanism for generating metric
perturbations that is completely universal, independent of the
particular realisation of inflation. The spectral properties of those
perturbations depends in a well known way on the precise dynamics
associated with a given model, but the mechanism is a generic property
of quantum field theory in curved space-times. What is less known is
that backreaction of the metric plus inflaton fluctuations on the
background space-time makes the inflaton field follow a Brownian
motion in which half of the time the inflaton field in a given domain
will jump upwards, instead of drifting down the effective
potential. In those domains the rate of expansion increases. Since the
amplitude of the quantum fluctuation is proportional to the local rate
of expansion, the inflaton field in a few domains may continue to jump
upwards, driving higher and higher rates of expansion, and therefore
those domains will eventually occupy larger and larger volumes. Thus
the universe becomes divided into domains in which the inflaton has
drifted down the potential towards reheating and others that are still
inflating, separated by distances much greater than their causal
horizon. This stochastic picture leads to the self-reproduction of the
universe and to eternal inflation. We may happen to be living in an
island of warmth appropriate for life in an otherwise cold and
eternally expanding universe. The island is probably large enough that
deviations from homogeneity and flatness are negligible up to many
times our present Hubble volume, so there is little chance of
observing those inhomogeneous regions, at least in our lifetime as
species. However, from the epistemological point of view, it is
fascinating to entertain the idea that in causally disconnected
regions of the universe we may have independent islands of thermalised
regions in all possible stages of their evolution.  Eventually all
causal domains will end inflation, but at any given time the bulk of
the universe is in eternal inflation. From this point of view, there
is not just one Big Bang, but infinitely many, corresponding to those
events where inflation ends locally and a radiation epoch ensues. This
alleviates the problem of initial conditions of inflation: once a
given domain allows a tiny burst of inflation, the flame will carry
away expanding the rest of the universe forever. Local regions of the
universe may collapse after matter domination, while others will
expand for ever. Our particular patch, what we call the observable
universe, is but a tiny insignificant spot in this metauniverse that
encompasses all of space and time. Most probably, this grand picture
of the universe will never be amenable to direct experimentation. We
may have a hint of its existence from the actual pattern of CMB
anisotropies, since the same quantum fluctuations that gave rise to
those inhomogeneities is also responsible for the ultra large scale
structure of the universe, but we may not be able to disprove it with
local observations.

\section{Variations on a common theme}

Although the paradigm of inflation is based on the simple idea that
the early universe went through a quasi-exponential expansion that
stretched all scales, there are in principle many ways to realise this
paradigm.  Our first reaction is, like Newton, to make no hypothesis,
but state that {\em everything works as if} an approximately constant
energy density dominated the evolution of the early universe.
Usually, this slowly changing energy density is parametrised by a
single scalar degree of freedom because it is the simplest
possibility, but for no other reason. This is enough to explain all
observations, both in the CMB anisotropies and in the LSS distribution
of matter.  However, one can think of a series of variations on the
same theme, for example adding another scalar field $-$ most high
energy physics models have not just one but many light scalar degrees
of freedom $-$ and induce isocurvature as well as adiabatic
perturbations. They would produce specific signatures in the CMB
anisotropies that have not been observed, and thus multi-field
inflation is essentially ruled out, at least to some degree.

It could also happen that the inflaton field may not be treated as a
free field because of its own self-coupling, because it couples to other
light fields, or due to its coupling to gravity. In either case, the
spectrum of quantum fluctuations is expected to have some, but small,
non-gaussian statistics (only the ground state of a free field is
actually a gaussian random field).  For the moment, these
non-gaussianities have been searched for, both in the CMB and in LSS,
without success. They would appear as deviations from the gaussian
expectations in the three- and four-point correlation functions of
galaxies and CMB anisotropies. A small degree of non-gaussianity is
expected at galactic scales due to non-linear gravitational collapse,
but for the moment what is observed is compatible with an evolved
primordial gaussian spectrum. However, in the future, the information
encoded in the non-gaussianities could be crucial in order to
distinguish between different models.

On the other hand, a prediction that is quite robust for slow-roll
inflation of the power-law type ({\em i.e.} with cubic and quartic
self-interactions of the inflaton) is a precise relation between the
deviations from scale invariance (the tilt of the scalar spectrum) and
the duration of inflation, which makes this tilt be negative, and of
order a few parts in a hundred. Another prediction of this class of
models is that the tilt is also scale-dependent, but the dependence is
very weak, of only a few parts in 10 000. These two predictions break
down in models that are non-slow-roll, which are compatible with the
rest of the inflationary phenomenology, but which predict large tilts
for both scalar and tensor spectra, as well as a large scale-dependence
of those tilts. Since the tilt has not been measured yet with sufficient
accuracy, we cannot rule out the possibility of non-slow-roll inflation.
Moreover, a large negative tilt would signal the presence of a large
tensor (gravitational wave) component which would then be seen in the
future CMB satellite experiments. However, this prediction is not
universal.  There are models of inflation, like hybrid models $-$ that
end inflation when the inflaton triggers the breaking of a symmetry
associated with another field to which it couples $-$ that can give an
essentially flat spectrum, {\em i.e.} no tilt within the precision of
planned missions, and also no scale-dendence of the tilt, nor a
significant tensor component.

\section{How to rule out inflation}

This is a tricky business since inflation is a paradigm with many
realisations. Inflation makes space flat and homogeneous, but at the
same time imprints space with metric fluctuations that later will seed
the local structures we observe, like galaxies and clusters of galaxies.
An obvious way to rule out the general idea of inflation would be to
find evidence of a large spatial curvature, but the question is on
what scale. Since the curvature of space depends on the duration of
the period of inflation $-$ the longer the flatter $-$ a relatively
short period might give rise to a marginally open or closed universe,
depending on what sign of curvature they started with. However, curved
models of inflation are somewhat fine-tuned and are nowadays ruled out
at a high confidence level by the position of the first acoustic peak
of the CMB temperature anisotropies.

Another way to rule inflation out is to discover that the universe is
topologically non-trivial because, as far as we know from the
classical theory of general relativity, the topology of space-time is
invariant under the expansion of the universe; the only thing
inflation does to topology is to increase the size of the topological
cell. For topology to be seen today as correlated patterns in the sky,
the cell should have a size smaller than the present Hubble volume.
Such patterns have not been seen by the WMAP precision measurements of
CMB anisotropies and, therefore, the topological cell, if it exists,
must have a size much bigger than our present horizon.  Yet another
way to rule out the general idea of inflation would be to discover a
global rotation of the universe, that is, a priviledged axis in the
sky. Since inflation expands all scales isotropically, such an axis
would stand out very clearly in the CMB sky today, and nothing like it
has been observed by WMAP.

The previous discussion has to do with the global background
properties of our observable universe. However, what characterises
inflation, and allows alternative models to be distinguished, is the
predicted spectrum of metric perturbations. Since the source of metric
perturbations is a fundamental or an effective scalar field (it could
be a condensate of some other field), there can be only two types of
perturbations at first order: scalar and tensor metric perturbations.
Vector perturbations only appear at second order, and therefore are
suppressed, unless there are peculiar additional sources acting during
inflation. Thus, a generic prediction of inflation is that the vector
perturbation spectrum is absent. If someone would measure a
significant primordial spectrum of vector perturbations in the CMB or
LSS, one would have to question some of the assumptions of inflation,
in particular single-field inflation, although it would probably not
rule out the idea altogether. For the moment we have not seen any
evidence of a vector perturbation in the CMB. This brings together an
immediate consequence with respect to the polarization spectra, as I
explained above. In the absence of any vector perturbation, parity is
conserved and the cross-correlations $\langle{\rm BT}\rangle$ and
$\langle{\rm BE}\rangle$ are expected to be zero. But this is a null
test, and it may happen that detailed observations of the CMB indicate
a non-zero primordial component. In that case, we would have to go
back to the drawing board.

Similarly with large tensor perturbations or gravitational waves. If
their amplitude is too large, compared with the scalar perturbations, it
would imply an origin of inflation close to the Planckian era, and thus
the classical description of space-time in terms of the general theory
of relativity would not be appropriate. There are models of inflation
that consider initial conditions close to the Planck boundary, but their
effects are seen on scales much much larger than our Hubble volume, and
by the time the fluctuations that gave rise to structure in our
observable universe were produced, inflation proceeded well below Planck
scale.

Concentrating now on the primordial scalar spectrum, the tilt of the
spectrum is not arbitrary. It cannot be too different from
scale-invariant ($n=1$), or inflation would not have occurred in the
first place. A deviation of order 100\% larger or smaller than 1 is
probably incompatible with most realisations of inflation, since it
would require too steep potentials, although there is no general
theorem, as far as I know. Typical slow-roll models of inflation
predict spectral tilts that are only a few percent away from scale
invariance.  Some models, like hybrid inflation, are perfectly
compatible with exact scale invariance, within experimental errors. In
order to get significant tilts one has to go beyond slow-roll, and
there are no generic analytical results for these models. One has to
study them case by case.

Another key feature of primordial spectra is the scale-dependence of
their spectral tilts. Again, most slow-roll models predict no
significant scale-dependence, just of order a few parts in 10 000. But
there could be specific models with particular features in their
effective potential that change significantly at scales corresponding to
those seen in the CMB and LSS, which would induce a scale-dependent tilt.
In fact, there has been some debate whether WMAP has actually seen such
a scale dependence in the CMB anisotropies, but there is no consensus.

In single-field models of inflation, the scalar and tensor spectra are
both computed from a single analytical function, the effective scalar
potential, and therefore there must a be a functional relationship
between these two spectra. In slow-roll inflation, it can be cast into
a relation between the ratio of the tensor to scalar amplitude and the
tensor spectral tilt. Such a relation is very concrete and generic to
all single-field models. We know of no other mechanism but inflation
which might actually give such a peculiar relation, and for that
reason it has been named ``the consistency relation'' of inflation.
If, once the gravitational wave spectrum is discovered, such a
relation did not hold, it would signal the demise of slow-roll
single-field models. However, deviation from that exact relation could
be due to multiple-field models, which predict a similar but different
relation, with one extra parameter, that can also be measured and then
checked to hold. If that doesn't happen either, then probably
slow-roll models are not a viable description, and we would need to do
fully non-perturbative calculations to find the exact relation between
the two spectra. Eventually, whatever the relation, there must be some
correlation between them that will allow cosmologists to discard all
but a few realizations of inflation.  That could be hard but fruitful
work.

The problem, however, could be observational: the foregrounds to the
CMB. In order to measure the tensor spectrum we need to detect the
B-component of polarization, which may be swamped by contamination
from a distorted E-component due to gravitational lensing in the
intervening mass distribution between us and the last scattering
surface. Moreover, on small scales it may be difficult to disentangle
the primordial spectrum from the observed spectrum of
perturbations. For instance, we know that non-linear gravitational
collapse induces both vorticity and non-gaussianities and that
gravitational lensing induces BE cross-correlation. In order to
disentangle foregrounds from the CMB anisotropies, we may have to know
more about the local and global distribution of matter in structures
like galaxies, clusters and superclusters.

Finally, most realizations of inflation describe the universe dynamics
with an effective scalar field that is essentially free during
inflation, only its mass is relevant. In that case, the ground state of
this quantum field gives rise to a gaussian spectrum of metric
fluctuations, with only second order deviations from gaussianity. Such a
prediction is very robust, at least in single-field models of inflation,
and therefore, if a significant non-gaussianity were observed in the CMB
anisotropies or LSS, it could signal multiple-field inflation, which
could be contrasted with other predictions like isocurvature signatures;
or it could indicate a significant contribution from postulated
modifications of gravity on large scales. In any case, ruling out
specific models of inflation leave fewer possibilities and allow
cosmologists to close in on the best model, which may eventually lead to
a final theory of inflation.

\section{Honest criticisms and false claims}

There are some features of inflation that are directly related with
its status as a paradigm, {\em i.e.} not yet a theory. The main
criticism that inflation has received over the years is that there is
no unique model, but a variety of single- and multi-field models, some
with fundamental descriptions, others purely phenomenological. I think
this is an honest criticism; in the absence of a fundamental theory of
inflation we have to deal with its effective realisation in terms of
specific models. However, the models are fully predictive, they give
concrete values for all observables, and can be ruled out one by one.
It is possible that in the near future, with more precise cosmological
observations, we may end up with a narrow range of models and
eventually only one agreeing with all observations.

There are, howover, some basic properties of inflation we still do not
comprehend. In particular, we ignore the energy scale at which inflation
occurred. It has often been claimed to be associated with the scale of
grand unification, in which all fundamental forces but gravity unified
into a single one, at around $10^{16}$ GeV, but this is still
uncertain. The only observable with direct information about the scale
of inflation is the amplitude of the gravitational wave spectrum, but we
have not measured that yet and it may happen that it is too low for such
a spectrum to be seen in the forseable future. This would be a pity, of
course, but it would open the door to low scale models of inflation and
a rethinking of the finetuning problems associated with them. A related
criticism is that of not knowing whether the inflaton field is a low
energy effective description of a more fundamental interaction or a
genuine scalar field. We still have not measured any fundamental scalar
field, and it could be that the inflaton is actually a geometrical
degree of freedom related to the compactification of some extra
dimensions, as recently proposed in the context of some string theory
realizations of inflation.

Moreover, the issue of reheating of the universe remains somewhat
speculative, even though a lot can be said in generic cases. For
instance, there are low energy models of hybrid inflation that could
have taken place $-$ albeit with some finetuning $-$ at the electroweak
scale. In that case, it is expected that the inflaton will be some
scalar degree of freedom associated with electroweak symmetry breaking
and thus in the same sector as the Higgs. The gravitational wave
component of the spectrum would be negligible, but at least one could
explore the reheating mechanism by determining the couplings of the
inflaton to other fields directly at high energy particle accelerators
like the Large Hadron Collider at CERN.

On the other hand, inflation has also been for many years the target
of unfounded criticisms. For instance, it has been claimed for years
that inflation requires extremely small (finetuned) couplings in order
to agree with CMB observations. Although this is true for some models,
this is not a generic feature of inflation: hybrid inflation models
agree very well with observations and do not require unnaturally small
couplings and masses. Another common error is to claim that inflation
is not predictive because of its many models. This is false, each
especific model makes definite predictions that can be used to rule
the model out. In fact, since the idea was proposed in the 1980's, the
whole set of open inflation models has been ruled out, as well as
models within a scalar-tensor theory of gravity. Furthermore,
multi-field models of inflation have been recently contrained by
observations of the temperature and polarization spectra, which do not
seem to allow a significant isocurvature component to mix with the
adiabatic one.

Another false claim which is heard from time to time is that inflation
always predicts a red tilt, with a smaller amplitude at smaller
wavelengths, but this is not true. Again, hybrid models of inflation,
which are based on the idea of spontaneous symmetry breaking in
particle physics, predict a slightly blue-tilted spectrum, and
sometimes no tilt at all. Associated with this claim, there was the
standard lore that the ratio of the tensor to scalar component of the
spectrum was proportional to the tilt of the scalar spectrum. This is
incorrect even in slow-roll models of inflation, being proportional to
the tensor tilt, not the scalar one, which makes the prediction of the
tensor amplitude more difficult.

\section{What inflation cannot answer}

Although inflation answers many fundamental questions of modern
cosmology, and also serves as the arena where further questions can be
posed, it is by no means the panacea of all problems in cosmology.  In
particular, inflation was not invented for solving, and probably cannot
solve by itself, the problem of the cosmological constant. This problem
is really a fundamental one, having to do with basic notions of gravity
and quantum physics. It may happen that its resolution will give us a
hint to the embedding of inflation in a larger theory, but I don't think
inflation should be asked responsible for not explaining the
cosmological constant problem. The same applies to the present value of
cosmological parameters like the baryon fraction, or the dark matter
content of the universe, or the rate of expansion today. Those are
parameters that depend on the matter/energy content after reheating and their
evolution during the subsequent hot Big Bang eras. At the moment, the
only thing we know for sure is that we are made of baryons, which
contribute only 4\% of the energy content of the universe. The origin of
the dark matter and the cosmological constant is a mystery. In fact, the
only thing we know is that dark matter collapses gravitationally and
forms structure, while vacuum energy acts like a cosmic repulsion,
separating these structures at an accelerated rate. They could both be
related to modifications of gravity on large scales $-$ associated with the
mysterious nature of the vacuum of quantum field theories $-$ but for
the moment this is just an educated guess.

Inflation can neither predict the present age of the (observable)
universe, nor its fate. If the present observation of the acceleration
of the universe is due to a slowly decaying cosmological constant, the
universe may eventually recollapse. Nor can inflation predict the initial
conditions of the universe, which gave rise to our observable
patch. Quantum gravity is probably behind the origin of the inflationary
domain that started everything, but we still do not have a credible
quantum measure for the distribution of those domains.

\section{Conclusions and Outlook}

The paradigm of inflation is based on the elegantly simple idea that the
early universe went through a quasi-exponential expansion that stretched
all scales, thus making space flat and homogeneous. As a consequence,
microscopic quantum field fluctuations are also stretched to cosmological 
scales and become classical metric perturbations that seed large scale
structure and can be seen as temperature anisotropies in the CMB.

The main predictions of this paradigm were drawn at least a decade
before observations of the microwave background anisotropies and the
matter distribution in galaxies and clusters were precise enough to
confirm the general paradigm. We are at the moment improving our
measurements of the primordial spectrum of metric fluctuations, through
CMB temperature and polarization anisotropies, as well as the LSS
distribution of matter. Eventually, we will have a direct probe of the
energy scale at which inflation took place, which may open the
possibility of constructing a complete theory of inflation.

For the moment, all cosmological observations are consistent with the
predictions of inflation, from the background space-time to the metric
and matter fluctuations. It is remarkable that such a simple idea should
work so well, and we still do not know why. From the epistemological
point of view, one could say that we are in a situation similar to
Newtonian mechanics in the nineteenth century, before its embedding in
Einstein's theory of general relativity.  We can predict the outcome of
most cosmological observations within the standard model of cosmology,
but we do not yet have a fundamental theory of inflation. 

Such a theory will probably require some knowledge of quantum
gravity. This has led many theoretical cosmologists into speculations
about realisations of inflation in the context of string theory or
M-theory. This is certainly an avenue worth pursuing but, for the
moment, their predictions cannot be distinguished from those given by
other models of inflation. Eventually the higher accuracy of future CMB
measurements and LSS observations will allow cosmologists to discard
most models of inflation but a few, in the search for the final theory
of inflation. It may require completely new ideas and perhaps even an
epistemological revolution, but I am sure that, eventually, we will be
able to construct a consistent theory of inflation based on quantum
gravity and high energy particle physics.

\section*{Acknowledgments}
\addcontentsline{toc}{section}{Acknowledgements}

It is always a pleasure to thank Andrei Linde for many enlightening
discussions on the paradigm of inflation and its far reaching
consequences. This work was supported by the Spanish Ministry of Science
and Technology under contract FPA-2003-04597.

\newpage

{\bf Juan Garc\'{\i}a-Bellido}

\begin{figure}[th]		
\centerline{\psfig{file=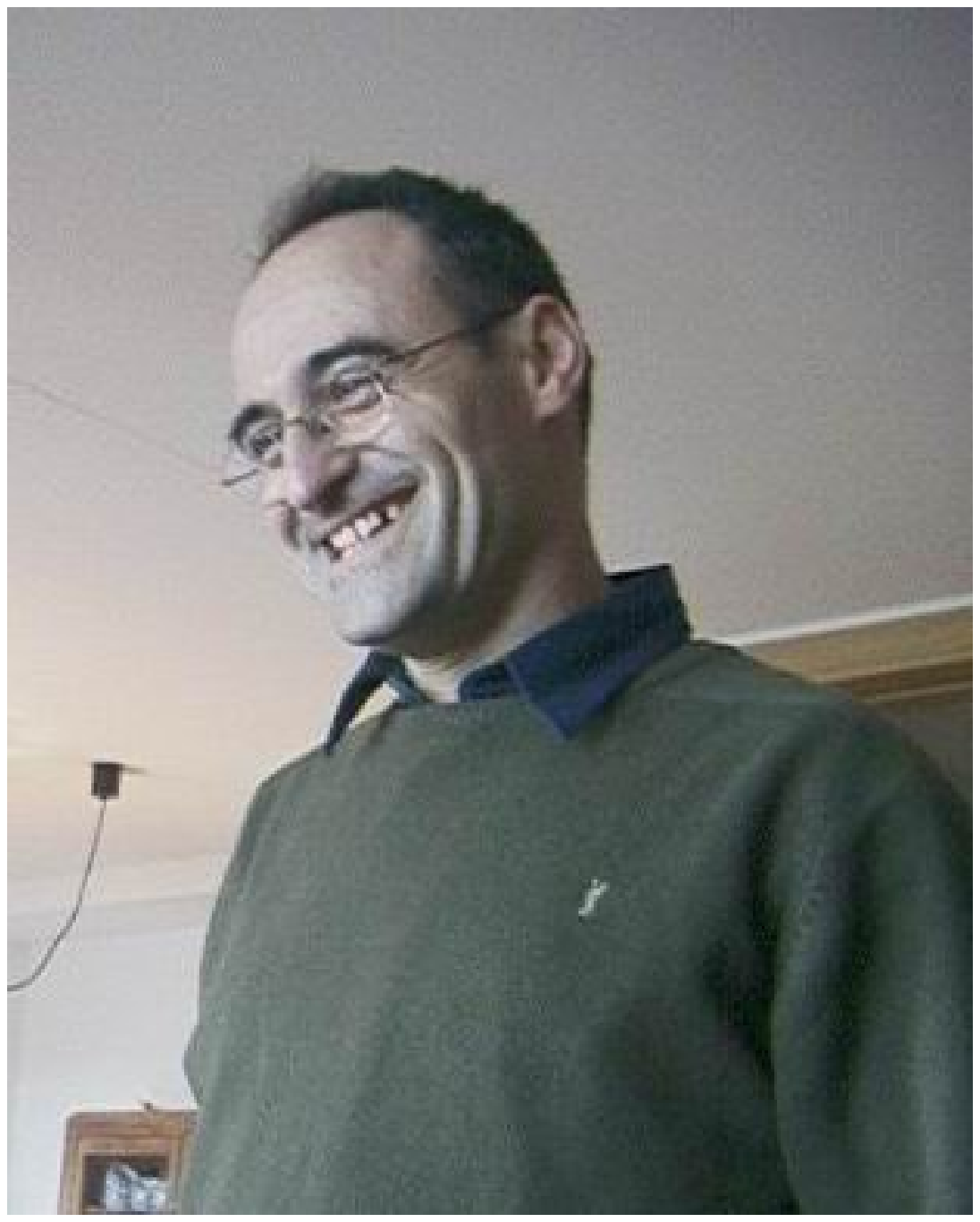,width=3in,angle=0}}
\vspace*{8pt}
\end{figure}

Born in 1966 in Madrid, Spain, Juan Garc{\'\i}a-Bellido obtained his PhD
in 1992 from the Aut\'onoma University of Madrid. He was a Postdoctoral
Fellow at Stanford University (1992-94), PPARC Fellow of the Astronomy
Centre at the University of Sussex (1995-96), Fellow of the TH-Division
at CERN (1996-98), a Royal Society University Research Fellow at
Imperial College (1998-99), and since then Professor at the Institute of
Theoretical Physics in Madrid. He has published around 70 papers in
theoretical physics and cosmology;  attended, organised and gave lectures
at international conferences and summer schools around the world, and is
referee of the most prestigious journals in the field. Aged 38, he is
married to a particle physicist and has two kids, a girl and a boy, aged
7 and 2, respectively.  Scientific interests include the early universe,
black holes and quantum gravity.

\end{document}